\documentclass[usenatbib,iop,numberedappendix]{aeb_emulateapj_2015}
\usepackage{amsmath}
\usepackage{amssymb}
\usepackage{xspace}
\usepackage[normalem]{ulem}
\usepackage{pifont}

\usepackage{natbib}

\usepackage[usenames,dvipsnames]{color}
\usepackage{hyperref}
\definecolor{darkblue}{rgb}{0.0,0.0,0.3}
\hypersetup{colorlinks,breaklinks,linkcolor=darkblue,urlcolor=darkblue,anchorcolor=darkblue,citecolor=darkblue}

\def\s{{\rm s}} 
\def\yr{{\rm yr}} 
\def\Gyr{{\rm G}\yr} 

\def\m{{\rm m}} 
\def\km{{\rm k}\m} 
\def\pc{{\rm pc}} 
\def\Mpc{{\rm M}\pc} 

\def\erg{{\rm erg}} 

\def\muas{\mu{\rm as}} 

\begin{document}

\title{
Impact of Accretion Flow Dynamics on Gas-dynamical Black Hole Mass Estimates
}

\newcommand{\perimeter}{3}
\newcommand{\waterloo}{1}
\newcommand{\wcfa}{2}

\author{
  Britton~Jeter\altaffilmark{\waterloo,\wcfa},
  Avery~E.~Broderick\altaffilmark{\waterloo,\wcfa,\perimeter,},
  B.~R.~McNamara\altaffilmark{\waterloo,\wcfa\perimeter,}
}
\altaffiltext{\waterloo}{Department of Physics and Astronomy, University of Waterloo, 200 University Avenue West, Waterloo, ON N2L 3G1, Canada}
\altaffiltext{\wcfa}{Waterloo Centre for Astrophysics, University of Waterloo, Waterloo, ON N2L 3G1, Canada}
\altaffiltext{\perimeter}{Perimeter Institute for Theoretical Physics, 31 Caroline Street North, Waterloo, ON N2L 2Y5, Canada}

\shorttitle{Gas Dynamics Systematics}
\shortauthors{Jeter et al.}

\begin{abstract}
At low redshift, the majority of supermassive black hole (SMBH) mass estimates are obtained from modeling stellar kinematics or ionized gas dynamics in the vicinity of the galaxy nucleus.  For large early type galaxies, stellar kinematics models predict higher masses than gas-dynamical models. In the case of M87, this discrepancy is larger than 2 $\sigma$.  Critical to gas-dynamical modeling is the assumed underlying dynamical state of the gas: that it lies on circular Keplerian orbits, potentially with some additional turbulent pressure support.  This is inconsistent with models of the gas flow about low-accretion-rate SMBHs and at odds with observations of the Galactic Center.  We present a simple model for non-Keplerian gas disks and explore their implications for SMBH mass measurements.  We show that a larger central black hole with gas experiencing small amounts of sub-Keplerian motion can produce velocity curves similar to models that just contain circular Keplerian motions and a lower black hole mass.  However, these non-Keplerian models are distinguishable from low-mass Keplerian models primarily through measurements of the velocity dispersion, wherein non-Keplerian models produce higher and narrower peak dispersions.  Away from the galaxy center, but still within the circumnuclear gas disk, non-Keplerian models also become distinguishable from Keplerian models via a shift in the velocity curve.  The velocity model presented in this paper is capable of resolving the discrepancy between the ionized gas dynamics and stellar kinematics mass estimates, and is applicable to gas-dynamical mass estimates of SMBHs in general. 
\end{abstract}


\section{Introduction}

Since first being implicated as the engines of quasars \citep{Lyn:1978}, it now appears that every luminous galaxy contains a supermassive black hole (SMBH), with masses ranging from $10^6M_\odot$ to $10^{10}M_\odot$, at its center \citep{Ferr-Ford:05}.
Despite composing less than 0.01\% of their host galaxy's mass, these are far from incidental; active galactic nuclei (AGNs), the rapidly accreting manifestation of SMBHs, produce enormous electromagnetic and kinetic luminosities, with $\approx 10\%$ driving relativistic jets that extend intergalactic distances.  SMBH masses are correlated with their host galaxy properties, including the velocity dispersion of bulge stars $\sigma$ \citep{Ferr-Merr:00, Geb-Msigma:00}, and total luminosity of the bulge \citep{Korm-Rich:95, Marc-Hunt:03}, and subsequent correlations with halo mass \citep{Korm-Bend:11}, and even the total number of globular clusters \citep{Burk-Trem:10, HandH:11}.  These are believed to arise due to feedback mechanism associated with the AGN phase, through which SMBHs play a critical role in regulating the growth and evolution of their hosts \citep[see, e.g.,][]{McNam-Nul:12, Fab:12}.  Nevertheless, the origin and impact of SMBHs remains only poorly understood, in part due to the difficulty in measuring SMBH masses.

Accurately estimating the SMBH masses is difficult due to the small spheres of influence, the region where the SMBH dominates the gravitational potential, and complicated nature of galactic centers.  A number of methods have been employed, with varying breadth of application and measurement precision.
These include applications of the Eddington limit \citep{Maz-Ban:17}, to the inversion of the above scaling relations \citep{Ferr:02}, and line reverberation mapping \citep{Bland-McKee:82, PeterReverb:11, ShenReverb:13}, all of which provide rough estimates of SMBH masses subject to various assumptions regarding AGN, their environments, and their relationships to their hosts.  More precise measurements, which provide the foundation for the empirical methods mentioned above, are obtained by modeling the motion of stars or gas in the nuclear region of the host galaxy \citep[see, e.g., the review by][]{Korm-Ho:13}.
In both cases, the underlying assumption is that relevant emitters are probes of the local gravitational potential, and thus when well within the sphere of influence of the central SMBH, its mass.  Characterizing gas motion around the central SMBH is critical to understanding how these objects accrete, and in turn how these accreting engines feedback on their host galaxies.

For the stellar-dynamical mass measurements, it is the velocity distribution of the stars that is employed as the probe of the gravitational potential.  In this case, the stars are presumed to be on ballistic orbits, determined by a gravitational potential set by both the central SMBH and the nuclear star cluster.  Typically, individual stars are not resolved, admitting only measurements of integrated line shapes.  Thus, based on the distribution of light, average spectra, and line shapes, the distribution of stellar masses, orbits, and mass distribution is reconstructed \citep{Korm-Rich:95, Gebh-Rich:00, M87stars:11}.  In contrast, for the gas-dynamical mass measurements, it is the velocity distribution of the orbiting ionized or molecular gas that is employed.  Typically, it is assumed that the gas is confined to a thin disk on circular Keplerian orbits, where the enclosed mass is estimated by directly applying Kepler's Law.  In many cases, while the line-of-sight velocity profile may be well described by a thin Keplerian disk, the velocity dispersions are not; the observed velocity dispersions can be up to an order of magnitude larger than predicted by the Keplerian disk model.  There have been attempts to account for the extra dispersion via additional pressure corrections, but it is not clear that this is well justified -- in many cases the effective temperatures associated with the observed velocity dispersions is well in excess of that necessary to destroy the molecules responsible for the observed line emission \citep{Macc-M87gas:97, Barth-Sarz:01, Neu-CenAGas:07, Walsh:10}.

For a handful of objects, SMBH mass estimates have been obtained using both methods, affording an opportunity to directly compare them.  While these are generally consistent in a number of nearby cases \citep{Dav-Thom:06, Past-Marc:07, Neu-CenAGas:07, Cap-CenAStars:07}, in nearly half there is a systematic difference between the stellar-dynamical and gas-dynamical mass estimates \citep{Verdoes:02, deFranc:06, M87stars:11, Walsh:12, M87gas:13}, with the latter typically being significantly smaller.
In a recent case, M87, this difference is roughly a factor of two, with the stellar-dynamical modeling finding a mass of $6.6\pm 0.4 \times10^9 M_\odot$ while the gas-dynamical modeling finds $3.5^{+0.9}_{-0.7}  \times10^9 M_\odot$.  This has clear implications for the (in)efficiency of M87's jet and current millimeter wavelength Very Long Baseline Interferometry (mm-VLBI) observations which promise to resolve the putative horizon.  

Sagittarius A* (Sgr A*), the SMBH at the center of the Milky Way, provides an elucidating example.
It has the virtue of having the most accurately measured mass of any SMBH, $4.3\pm 0.3 \times10^6 M_\odot$, obtained via the observation of orbiting massive stars \citep{Gil-SgrAStars:09, GillS2:09, Ghez:09}.\footnote{This differs from stellar-dynamical mass measurement in that Sgr A* provides the only example for which individual stars may be resolved and tracked on decadal timescales.  The orbital of S2, one of the massive stars used for this purpose, passes within 120 au of the SMBH and has a period of 16 yr.}
It also has a variety of historical gas-dynamical mass measurements, the earliest of which had similar spatial resolution, measured in terms of the size of the sphere of influence of the central SMBH, as those recently reported of extragalactic SMBHs \citep{Lacy:79, Lacy:80}.  That is, by measuring the velocities and disperisions of lines emitted by ionized gas clouds, \citet{Lacy:80} estimated a central SMBH mass of $2.4\times10^6 M_\odot$, albeit with 100\% error.\footnote{While \citet{Lacy:80} reports a mass of $3\times10^6 M_\odot$, they assume that the Galactic center is at a distance of 10~kpc; we provide the value after correcting this distance to 8~kpc, consistent with the most recent measurements.}

For Sgr A*, the reason for the discrepancy is clear.  Subsequent observations of the Galactic center have fully resolved the ionized gas within the sphere of influence of the SMBH.  Contrary to the assumption in \citet{Lacy:80}, the gas does not move along circular, Keplerian orbits.  Rather, it is organized into the ``mini''-spiral, a pc-scale structure with multiple arms and distinct non-Keplerian motions \citep{Beck:82, Mont-Hern-Ho:09, IronsLacy-SgrASpiral:12}.  The origin of the structures in the Galactic center is the larger-scale, tri-axial Galactic potential, and its interaction with the molecular torus at 3~pc \citep{Eckart:02, Scho:02, Genz:10}.  When the non-Keplerian, non-circular gas structures are modeled, the revised gas-dynamical mass estimate is $4.5\times 10^6 M_\odot$, in agreement with that derived from stellar orbits \citep{IronsLacy-SgrASpiral:12}.

Even outside our own galaxy, when the nuclear gas disk is well resolved and gas velocity profile carefully mapped, the mass estimate from gas kinematics is entirely consistent with stellar dynamics mass estimates \citep{Davis_etal:17, Boiz-Barth-Walsh:19}. However, in galaxies where there is not a well-resolved disk, there are theoretical reasons to believe that, in sub-Eddington systems, inside the Bondi radius the gas does not lie on Keplerian or near-Keplerian orbits \citep{Nar-Yi-ADAF:94, Neu-CenAGas:07, Chan-Krol:17, Iman:18}.  Hence, here we explore the impact of sub-Keplerian velocity profiles on gas-dynamical mass estimates.

Many prior attempts to explore deviations from Keplerian motion have been made.  These typically invoke a turbulent effective pressure, $P_{\rm eff}=\rho \sigma^2$, within the gas disk \citep{Neu-CenAGas:07}.  In these, the modified orbital velocity is parameterized through the choice of $P_{\rm eff}$.  In no case is a significant radial velocity considered.

In contrast, we parameterize the velocity profile directly, motivated by radiatively inefficient accretion flow (RIAF) models \citep{Nar-Yi-ADAF:94, Blan-Begel:99, Chan-Krol:17}.  These occur when the mass accretion rate at the black hole falls below 1\% of the Eddington rate, and incorporate potentially substantial mass loss via winds, and describe the accretion flow inside the Bondi radius \citep{Park-Ostr:99}.  In the absence of fully specifying the gas disk structure, we parameterize the orbital and radial velocities directly
\begin{equation}
  v_r = - \alpha v_k
  \quad\text{and}\quad
  v_\phi = \Omega v_k
 \label{eqn:vrvphi}
\end{equation}
for constants $\alpha$ and $\Omega$.  For advection dominated accretion flow (ADAF) models, $\alpha\lesssim0.1$, and $\Omega\lesssim0.4$ \citep{Nar-Yi-ADAF:95, Nar-Mah-Quat:98}.  For RIAFs these can be more modest \citep{Quat-Nar:99}.  This two parameter model does directly not address the physical origin for the modified accretion flow; doing so would require modeling the global structure of the flow.  However, in principle, measuring $\alpha$ and $\Omega$ would provide a means to reconstruct the radial density and temperature profile.  Even modest deviations of $\Omega$ from unity impose large systematic uncertainties on the reconstructed SMBH mass.  Within the SMBH sphere of influence, assuming a circular Keplerian flow, the mass estimate is given by $M = r v_\phi^2/G = \Omega^2 M_{\rm true}$.  Thus, setting $\Omega=0.71$, consistent with RIAF models, would reduce the mass by a factor of 2.  Less obvious is that the introduction of a non-zero radial velocity has significant implications for the velocity dispersion.

We approach the discussion of the systematic impact on SMBH mass estimates via M87, due to the recent disagreement between gas- and stellar-dynamical measurements.  In doing so we follow closely the analysis of \citet{M87gas:13}, adopting elements of their observations and emission model.  However, it should be understood that our conclusions are applicable to gas-dynamical mass measurements generally.  In Section 2, we describe the disk model in detail and how emission lines are computed.  Implications for spatially resolved spectral observations are presented in Section 3.  In Section 4 we discuss the implications of our model for the mass of M87, and the implications for the $M$-$\sigma$ relationship generally.  We collect conclusions in Section 5.  We assume a distance to M87 of $16.7~\Mpc$ \citep{Bird-Harris:10, Cantiello:18}.  

\section{Modeling Central Gas Motion}

As mentioned above, we approximately reproduce the procedure for modeling emission lines described in \citet{M87gas:13}.  Since gas dynamics mass estimates are predicated on the observation of nuclear emission lines, our model simulates the observation of emission lines from gas flows that exhibit sub-Keplerian motion.  We assume the ionized gas lives in clouds with temperatures at or below $10^{4} K$, and the clouds themselves move in the potential of the central black hole on virialized orbits.  Once we model the intrinsic line shape for gas inside the clouds, we generate a parameterized cloud velocity field with separate radial and azimuthal components.  This intrinsic line is then boosted to produce the correct intensity for a far away observer.  Lastly, we light up our gas disk with a pair of emissivity profiles derived from observations of M87's nuclear region \citep{M87gas:13} to produce line intensities approximately consistent with observations.  This broadened line is then smeared in the image space with an elliptical Gaussian kernel simulating the resolution of the Hubble Space Telescope (HST) Space Telescope Imaging Spectrograph (STIS) instrument. 

For a gas emission line, the general line shape is a Voigt profile, a combination of Gaussian and Lorentzian components \citep{Padm-book:00}.  However, far from the galactic nucleus, the observed line widths are typically very narrow, implying that the processes responsible for the Lorentzian component (pressure broadening, natural line width) may be neglected \citep{Neu-CenAGas:07, M87gas:13}.  Thus, we model the natural line with a Gaussian profile
\begin{equation}
  \begin{gathered}
    \phi_{0}(\nu) = \frac{1}{\sqrt{2\pi \Delta\nu_D^2}} \exp \left[ -\frac{(\nu-\nu_0)^2}{2\Delta\nu_D^2} \right], \\
    \text{where}\quad
    \Delta\nu_D = \nu_0 \frac{\sigma}{c}
  \end{gathered}
  \label{eqn:line_width}
\end{equation}
where $\sigma$ is the velocity dispersion of the emitting gas, and $\nu$ and $\nu_0$ are the thermally broadened rest-frame line frequency and unbroadened line center, respectively.

The existence of line emission near the galactic nucleus implies the temperature of the emitting gas must be less than $T\approx10^4$~K, otherwise the emitting gas would be start to become fully ionized.  However, measured central dispersions on the order of $100 ~{\rm km s^{-1}}$ \citep{Neu-CenAGas:07, Walsh:10, M87gas:13} imply temperatures $>10^6 $~K if derived solely from thermal broadening, which should completely ionize the gas and preclude any line emission in the galactic nucleus.  Therefore, the existence of wide emission lines near the galactic nucleus motivates a picture in which cooler, line-emitting clouds are embedded in a large-scale, partially virialized flow.  In the limit of many such clouds, the resulting lines will be Gaussian with line widths dominated by the dispersion in the turbulent cloud velocities.

From the virial theorem, the line-of-sight dispersion velocity $\sigma$ is related to the gravitational potential energy via 
\begin{align}
 \sigma^2 &=  f \frac{GM}{R}. 
 \label{eqn:sig_virial}
\end{align}
where $M$ is the mass enclosed at a radius $R$, and $f$ is a numerical factor, typically of order unity for a purely virial, dispersion supported system.  For a thin gaseous disk, it is possible to relate the height to the temperature via $h=r (c_{s}/v_{k})$, where $c_s$ is the speed of sound.  For a monatomic ionized gas
\begin{align}
 (c_s)^2 &= \frac{5}{3}\frac{P}{\rho} = \frac{5}{3}\frac{kT}{\mu m_p}
 \label{eqn:disk_sound}
\end{align}
where $P$ is the pressure in the disk, $\rho$ is the density, $T$ is the gas temperature, $\mu=0.6$ is the mean molecular weight for an ionized disk, and $m_p$ is the proton mass. It is possible to define a virial temperature as $kT_{vir}\approx GMm_{p}/r$, and then define the disk scale height as
\begin{align}
 \frac{h}{r} &= \left(\frac{5}{3\mu} \right)^{\frac{1}{2}} \left(\frac{T}{T_{vir}} \right)^{\frac{1}{2}} 
 \approx 1.67 \left( \frac{T}{T_{vir}}\right)^{\frac{1}{2}}.
 \label{eqn:scale_height}
\end{align}
When a disk is turbulent, it is possible to use $T$ as an effective temperature to describe both thermal and turbulent contributions to the gas motion, and we can then replace $c_s$ with $\sigma$, giving 
\begin{align}
 \sigma^2 &= (v_k)^2 \left( \frac{h}{r} \right)^2 \sim \frac{GM}{r} \left( \frac{h}{r} \right)^2
 \label{eqn:disk_sig}
\end{align}
For very cold disks, $h/r$ can be small, but for a typical RIAF, $h/r \sim 0.3$ so $\sigma^2 \sim 0.1 GM/R$, or $f=0.1$.  

It is also possible to empirically estimate $f$ using the observations in \citet{M87gas:13} assuming the gas dispersion can be entirely associated with an effective temperature, with both turbulent and thermal components.  The minimum velocity dispersion is measured to be about $150 ~\km~\s^{-1}$ at $40~\pc$ from the center.  For a fully virial, dispersion supported system with the same central mass, this velocity dispersion should occur at a radial distance of $\approx ~400~\pc$.  We thus adopt an $f=0.1$ going forward, as the ratio of these distances is consistent with the predicted numerical factor for an RIAF type disk.

We model the the global gas cloud motion assuming the clouds lie in a thin disk around the central black hole, tilted at an inclination $i$ with respect to the observer's line of sight.  The velocity field, $\vec\beta$, of this disk is parameterized with a radial and azimuthal component 
\begin{equation}
  \vec{\beta} = \frac{v_r}{c} \hat{r} + \frac{v_\phi}{c} \hat{\phi}, \label{eqn:vel_field}
\end{equation}
where $v_r$ and $v_\phi$ are given in Equation (\ref{eqn:vrvphi}), and $\hat{r}$ and $\hat{\phi}$ are the radial and azimuthal unit vectors relative to the central black hole and aligned in the normal way with the disk axis.  We can rewrite this in terms of Cartesian coordinates $X$ and $Y$, defined such that the $Z$-axis is aligned with the disk axis, and the $X$-axis is parallel to the observer's $x$-axis:  
\begin{align}
\vec{\beta} &= \frac{v_r}{c}\left( \frac{X}{R} \hat{X} + \frac{Y}{R} \hat{Y}\right) 
+ \frac{v_\phi}{c} \left( -\frac{Y}{R} \hat{X} + \frac{X}{R} \hat{Y} \right) \nonumber \\
&= \left( v_r X - v_\phi Y \right) \frac{\hat{X}}{Rc} + \left( v_r Y + v_\phi X \right) \frac{\hat{Y}}{Rc}. \label{eqn:beta_cart}
\end{align}
For a distant observer with a line of sight, $\vec{k}=\cos i ~\hat{Z}+\sin i ~\hat{Y}$, the projected velocity, $\vec{k} \cdot \vec{\beta}$, is 
\begin{equation}
  \vec{k} \cdot \vec{\beta}
  = \frac{\sin i}{Rc} \left( - \alpha v_{k} Y + ~\Omega v_{k} X \right),
  \label{eqn:k_dot_beta}
\end{equation}
where we have expressed $v_r$ and $v_\phi$ in terms of $v_k$ as described in Equation (\ref{eqn:vrvphi}).  

We construct the observed line shape, including Doppler beaming and the Doppler shift, via the Lorentz invariant $I_\nu/\nu^3$:
\begin{equation}
  \phi(\nu)
  =
  g^{-3} \phi_0(g\nu)
\end{equation}
where $g$ is the standard Doppler factor,
\begin{equation}
  g = \frac{1-\vec{k} \cdot \vec{\beta}}{\sqrt{1-\beta^2}}.
  \label{eqn:doppler}
\end{equation}

Finally, we use an empirically motivated emissivity model following the prescription outlined in \citet{M87gas:13}, where the emissivity is modeled by fitting a number of Gaussian components to the observed light profile.  In that work, the observed light profile is fit best by two offset elliptical Gaussians, but for clarity we use a pair of concentric circular Gaussians, $j_1$ and $j_2$:
\begin{equation}
  \begin{aligned}
  j_{1} &= A\exp \left( \frac{-R^{2}}{2r_{1}^{2}} \right) \\
  j_{2} &= B\exp \left( \frac{-R^{2}}{2r_{2}^{2}} \right) \label{eqn:j2},
  \end{aligned}
\end{equation}
where $r_1$ and $r_2$ are the widths of the emissivity profiles, and $A$ and $B$ are numerical scaling factors.  For M87, we adopt the values $r_{1}=6.6~\pc$ and $r_{2}=23.7~\pc$, and $A=26.5$, $B=1.0$ in arbitrary flux units, borrowing from Table 1 in \citet{M87gas:13}.  The observed line intensity is then
\begin{equation}
I_{\nu} = \left( j_{1} + j_{2} \right) \phi(\nu)
\end{equation}

\begin{figure*}[!ht]
  \begin{center}
    \includegraphics[width=\textwidth]{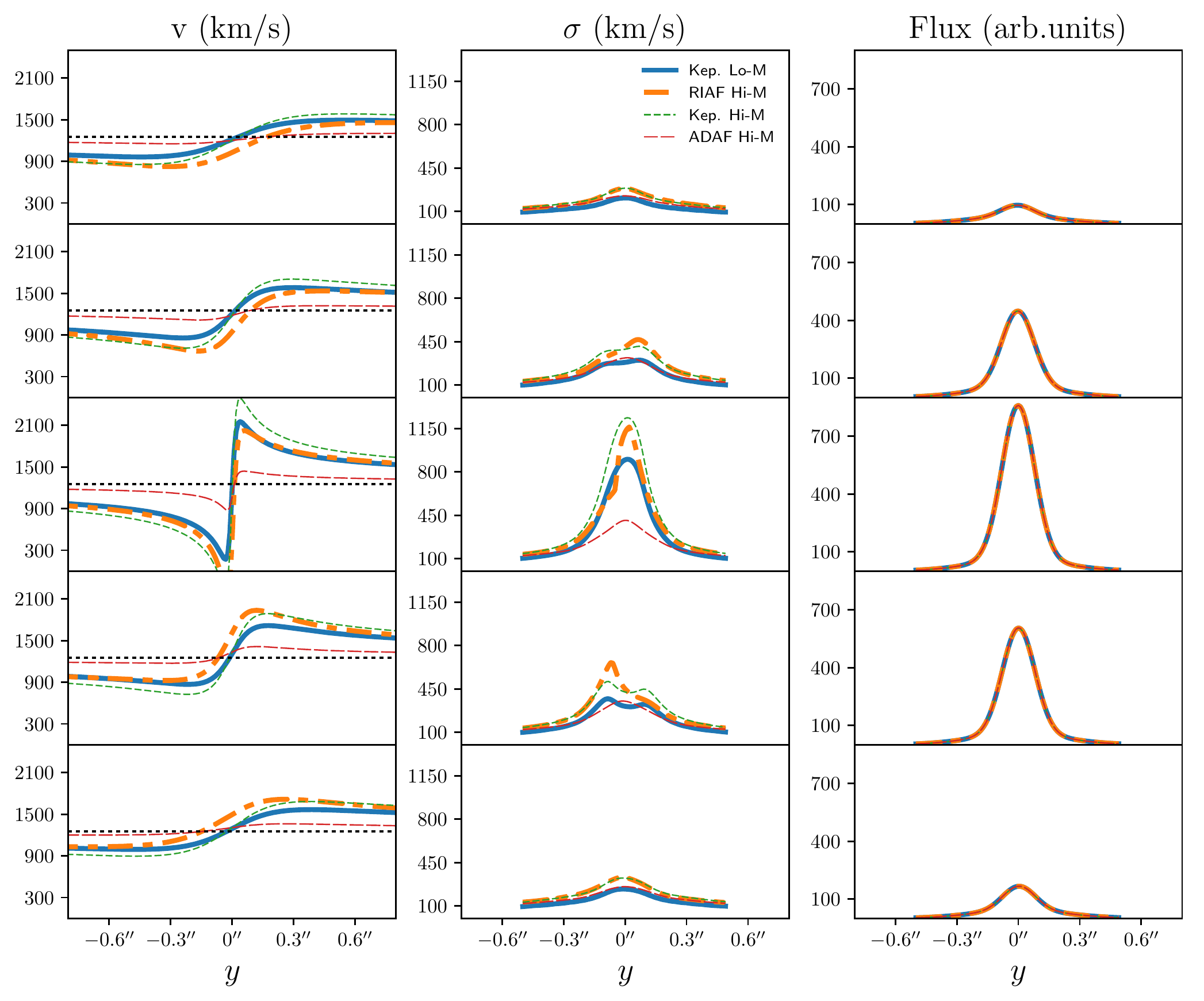}
  \end{center}
  \caption{Line-of-sight velocity (left), dispersion (middle), and integrated flux (right) along the slit vertical axis for five different horizontal slit positions.  From the top row, and centered on the black hole, the first slit is centered at $x=-0.2 ''$ , the second slit at $x=-0.1 ''$, the third slit at $x=0.0 ''$, the fourth slit at $x=0.1 ''$, and the fifth slit at $x=0.2 ''$.  The Keplerian low-mass model is is plotted as a thick solid blue line in all columns, the RIAF high mass model as a thick orange dashed-dotted line, the Keplerian high-mass model as a thin green dashed line, and the ADAF model as a thin red long-dashed line.}
\label{fig:KepCompare}
\end{figure*}

We produce a $2.42'' \times 2.42'' \times 393 \angstrom$ data cube representing $x$ and $y$ pixel coordinates and $\lambda$, respectively, in the image plane, with a $0.01''$ spatial resolution and a $1 ~\angstrom$ wavelength resolution centered on $6548 ~\angstrom$, the natural line frequency of [N II].  To project the image plane onto the disk, we counter-rotate the data cube by a position angle $\vartheta$ and then tilt by an inclination $i$:
\begin{align}
X &= x ~ \cos{\vartheta} - y~\sin{\vartheta}, \label{eqn:xrotate}\\
Y &= \frac{x ~\sin{\vartheta} + y ~\cos{\vartheta}}{\cos{i}} \label{eqn:yrotate}
\end{align}
where the $\vartheta = 6^\circ$ and $i = 42^\circ$ are again taken from measurements in \citet{M87gas:13}.  We simulate an HST spectrograph observation by convolving the data cube with a elliptical Gaussian kernel with standard deviations $\sigma_x = 0.1''$ and $\sigma_y = 0.0507''$, which has the effect of blurring the spectral features in the image plane.

\section{General Observational Implications}

In order to characterize any differences in velocity and dispersion profiles between Keplerian and non-Keplerian velocity flows, we simulate observations at five different x-locations, corresponding to slits, covering the inner $0.5'' \times 1.0''$ region centered on the black hole.  For each of these simulated observations, we extract the velocity and integrated flux along the $y$-direction, and construct the dispersion by using spline interpolation to find the FWHM of the blurred line profile.

\begin{figure*}[!ht]
  \begin{center}
    \includegraphics[width=\textwidth]{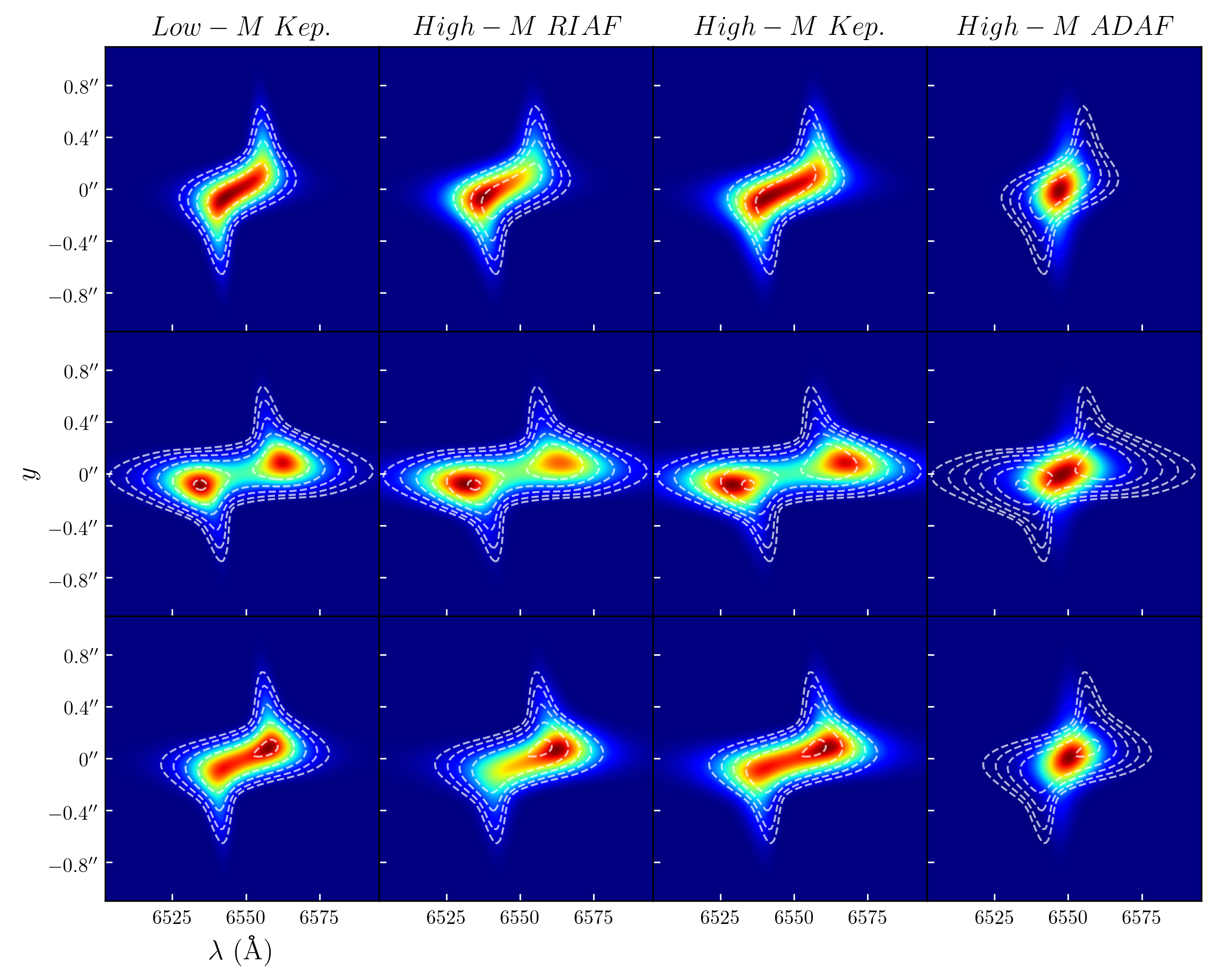}
  \end{center}
  \caption{2D spectra along the slit for a the case of a low-mass central black hole with a Keplerian velocity field (first column), a high-mass central black hole with a sub-Keplerian (RIAF) velocity field (second column), a high-mass black hole with a Keplerian velocity field (third column), and a high-mass central black hole with an ADAF velocity field (fourth column). The top and bottom rows correspond to slits centered $0.1''$ away from the central slit.  White dashed contours for the low-mass Keplerian model are shown in all panels. }
\label{fig:2dSpectra}
\end{figure*}

The only parameters that we vary between each model are the central black hole mass $M$ and the values for $\alpha$ and $\Omega$.  Our baseline model is an attempt to produce similar results to those in \citet{M87gas:13}, and has a black hole mass of $3.5\times10^9M_\odot$ and circular, Keplerian velocity field: $\alpha=0$ and $\Omega=1.0$.  We also produce a model with circular Keplerian velocities for the mass estimate from \citet{M87stars:11}, $M=6.6\times10^9M_\odot$, where $\alpha=0$ and $\Omega=1.0$.

We compare these to two non-Keplerian velocity profiles, assuming the high-mass estimate for the black hole mass in both.  Following \citet{Nar-Yi-ADAF:95}, we consider an ADAF-like velocity profile, with $\alpha=0.072$ and $\Omega=0.2$, arising from their self-similar model \citep[see Fig. 1 of][]{Nar-Yi-ADAF:95}.  We also consider a more modest sub-Keplerian flow, similar to RIAF models, for which we set $\alpha=\sqrt{0.1}$ and $\Omega=\sqrt{0.7}$.

We compare radial velocity, velocity dispersion, and flux along the $y$-direction for these four models in Figure \ref{fig:KepCompare}.  The Keplerian low-mass model represents the typical model for ionized gas motions in SMBH mass estimation experiments.  In this model, the gas has no radial velocity component, and the azimuthal velocity component is equal to the circular Keplerian velocity.  The velocity curves are approximately consistent with the model produced by \citet{M87gas:13}, and produce the expected symmetry across the slit center.  The dispersions produced in our model are different than in typical gas-dynamical modeling.

Unlike \citet{M87gas:13}, we assume a spatially variable turbulent dispersion, increasing with the virial temperature as described in the previous section.  This produces two notable effects in the dispersion plots: the first is a higher dispersion away from the slit center compared to the gas dynamical models presented, e.g., in \citet{M87gas:13}.  Second, there is a higher peak dispersion in the slit center.  The first effect almost entirely mitigates an observational discrepancy between the modeled and observed dispersions in \citet{M87gas:13} and others, where an additional $\sim 100 ~\km~\s^{-1}$ is added to the modeled constant dispersion to achieve closer fits to observed data.  The second effect only produces higher dispersions than other works in the center of the central slit, where the virial dispersion can reach as high as $900~\km~\s^{-1}$, much higher than the order $100~\km~\s^{-1}$ seen in constant-dispersion models.  Both of these effects are sensitive to the radial emission profile in the disk.

\subsection{Line-of-sight Velocities}

The high black hole mass Keplerian model produces qualitatively different velocity and dispersion profiles than the low-mass Keplerian model.  The velocity curves are still symmetric around the center, but all slits exhibit higher peak velocities, up to $200~\km~\s^{-1}$ higher than the low-mass Keplerian model in the central slit.  The ADAF model produces radial velocities that are dramatically suppressed compared to the low-mass Keplerian model, even though the black hole mass used for the ADAF model is the same as the high-mass Keplerian model.  The RIAF model is most similar to the low mass Keplerian model for the radial velocity curves, and in the central slit the velocity curves for the RIAF and low-mass Keplerian model effectively coincide.  In the other slits, the RIAF velocity curve looks like the low-mass Keplerian curve, but shifted approximately $0.1 ''$ right or left depending on whether the slit is to the left or right of the central slit.  The RIAF velocity profile is still similar to the low-mass Keplerian curve outside the inner $0.6 ''$ in all slits.  Due to typical observational uncertainties, a gas disk exhibiting even substantially non-Keplerian motions could be mistaken for a Keplerian gas disk with a smaller central black hole mass, given only the velocity data. 

\subsection{Velocity Dispersions}

\begin{figure*}[!ht]
  \begin{center}
    \includegraphics[width=\textwidth]{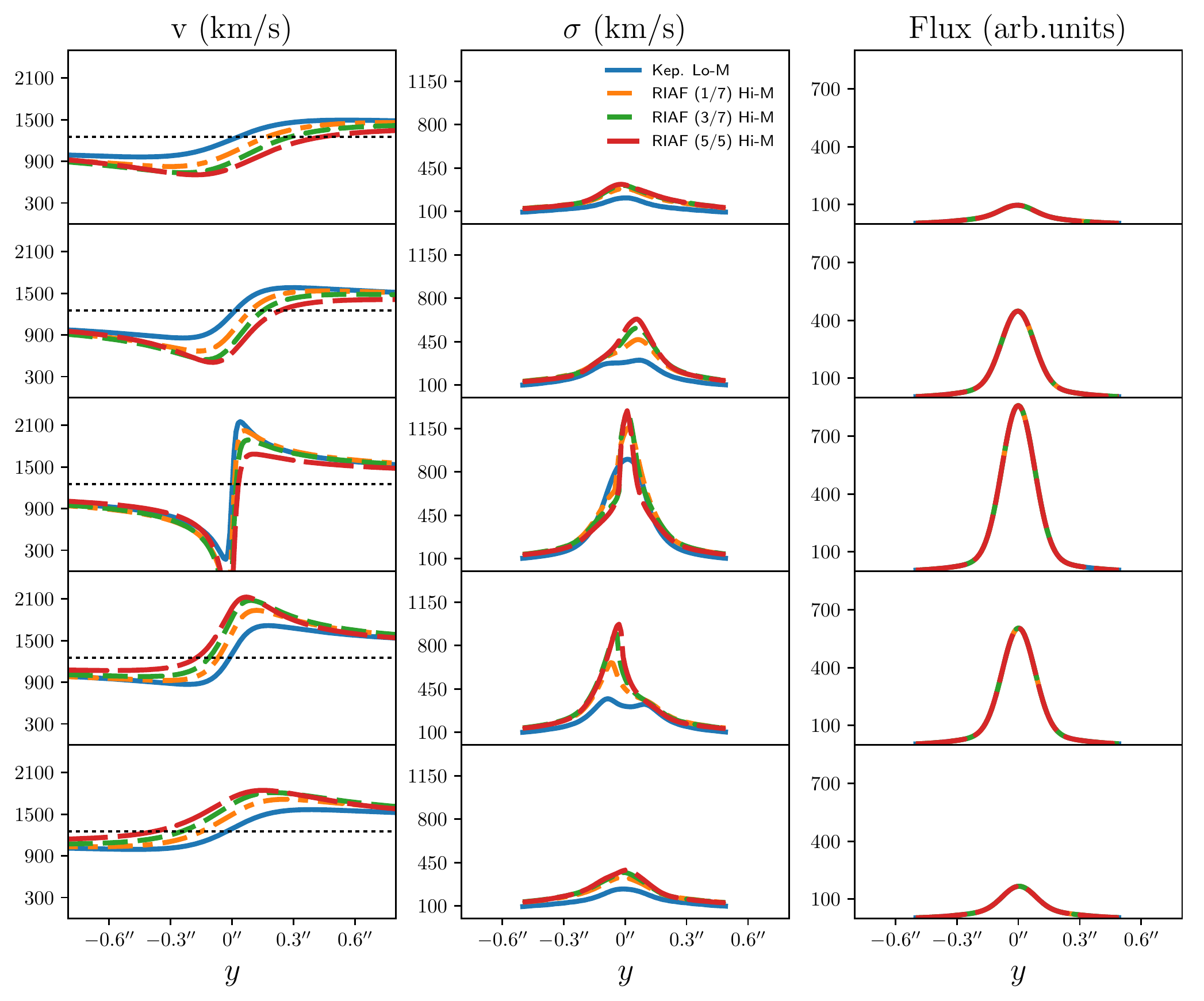}
  \end{center}
  \caption{Line-of-sight velocity (left), dispersion (middle), and integrated flux (right) along the slit vertical axis for five different horizontal slit positions.  The slit positions are the same as in Figure \ref{fig:KepCompare}.  The Keplerian low mass model is is plotted as a solid blue line in all columns, the slightly radial RIAF high-mass model as an orange dashed-dotted line, the RIAF model with additional radial motion as the green dashed line, and the RIAF model with even distribution of radial and azimuthal motion as the red long-dashed line}
\label{fig:RIAFCompare}
\end{figure*}

The RIAF model is distinguishable from the low-mass Keplerian model in the distribution and magnitudes of the velocity dispersions.  The RIAF model produces higher dispersions in the inner $0.6''$ than the low-mass Keplerian model, and even produces higher dispersions than the high-mass Keplerian model in outer slits.  In the central slit, the RIAF model produces peak dispersions very near the high-mass Keplerian model, roughly $300~\km~\s^{-1}$ higher than the low-mass Keplerian model.  Another significant feature of the RIAF dispersion profile is the narrower peak in the central slit.  Because of this, the RIAF model produces qualitatively the same dispersions as the low-mass Keplerian model outside of the inner $0.2''$.  While the peak dispersion is higher than the low-mass model, in practice distinguishing the RIAF and low-mass Keplerian model in the central slit may be difficult if the slit resolution is poor.  However, outside the central slit, the RIAF dispersions are generally higher than those associated with the low-mass Keplerian model, and exhibit asymmetries due to the radial gas motions, and are, therefore, capable of distinguishing between the two models.

The high mass Keplerian model produces dispersions in the central slit almost $400~\km~\s^{-1}$ higher than the low-mass Keplerian model, peaking near $1200 ~\km~\s^{-1}$.  The ADAF model produces velocity dispersions that are very similar to the low-mass Keplerian model in every slit except for the center slit.  In the center slit, the ADAF dispersion is lower than any other model by at least $300 ~\km~\s^{-1}$, and is qualitatively distinguishable from the low-mass Keplerian model.  This further supports the conclusion that spatially resolved velocity dispersions provide a key signature of non-Keplerian flow velocities.

\subsection{Spatially Resolved Spectra}

In Figure \ref{fig:2dSpectra} we produce spatially resolved spectra for the three innermost slits by projecting the emission line along the slit y-direction.  Dashed contours in this figure are emission intensities for the low-mass Keplerian model, and are plotted to facilitate comparisons with other models, and the rest-frame wavelength is $6548\angstrom$. 

In the central slit, the RIAF model produces a similar spectral profile to the low-mass Keplerian model, peaking approximately at the same frequency and producing the same intensity distribution across the slit.  The primary difference between the RIAF and low-mass Keplerian spectra is the asymmetry in intensity between the long and short wavelength peaks. In the Keplerian model, both peaks have approximately the same intensity, but long wavelength peak in the RIAF model is dimmer than the short wavelength peak, compared to both models.  When we compare the low-mass Keplerian and high-mass Keplerian models, we can see that the high-mass Keplerian model produces a wider emission profile in wavelength.  The emission peaks in the high-mass Keplerian model are less coincident in wavelength with the low-mass model than even the RIAF model.  The ADAF model is qualitatively different from the low-mass Keplerian model, and produces little to no emission outside $6525\angstrom$ and $6575\angstrom$.  

In the outer slits, the RIAF model still has a similar emission profile to the low-mass Keplerian model, but red- or blue-shifted approximately $10\angstrom$ depending on which side of the disk the slit observes, and the emission peaks preferentially in the same direction as the Doppler shift.  The high-mass Keplerian model again produces more emission across wavelengths, and the central emission ridge (orange and red regions in Figure (\ref{fig:2dSpectra})) is larger than in the low-mass Kelplerian model.  The ADAF model still produces a small emission region compared to all the other models, and the emission peak is Doppler shifted from the rest wavelength by much less than the other models.  

The line modeling done here is only for a single emission line.  The observations in \citet{M87gas:13} measure the H$\alpha$ and [NII] emission line complex, and the analysis produced in that work only incorporates data from spectra that had been satisfactorily decoupled.  Close to the center, the line dispersions become large enough such that the three emission lines begin to overlap, and uniquely decomposing the lines becomes difficult.  Since velocity and dispersion measurements in the very center are omitted in \citet{M87gas:13}, it is difficult to compare this work with the observational results, as the most significant differences occur where line-decomposition becomes most difficult.  The results of the modeling done in this work could exacerbate this issue; one of the key results of this work is increased dispersion for non-Keplerian models away from the center slit.  This increase can blend nearby emission lines further out from the center, and make decomposition more difficult.  

\subsection{Trends in Non-Keplerian Models}

As mentioned above, it is possible to distinguish the low-mass Keplerian model from the RIAF model via a shift in the RIAF velocity curves when observing away from the central slit, and via increased and narrower RIAF dispersion profiles compared to the low-mass Keplerian model.  To better characterize these differences, we produce radial velocity and dispersion curves for two more non-Keplerian models to explore the effects of increased radial velocities in Figure \ref{fig:RIAFCompare}.  These other two models both use the higher black hole mass, and have a total velocity magnitude equal to the high-mass Keplerian model.  The first of these new models has a radial velocity component of $\alpha=\sqrt{0.3}$ and an azimuthal velocity component of $\Omega=\sqrt{0.7}$.  Compared to our proto-typical RIAF model, this has an increased radial velocity component.  The other new model has $\alpha=\Omega=\sqrt{0.5}$.  This model has an even larger radial velocity component, in exchange for a reduced azimuthal contribution.

As we increase the radial velocity, the velocity curves shift further left or right in the y-direction depending on the side of the disk we measure the dispersed spectra on, and become more peaked on that side of the shift.  This shift can be as large as $0.3''$ for observations at the edges of the disk, but this shift is suppressed near the galaxy's center.  In the center, increasing the radial velocity component produces a shift down in velocities, where the difference in peak velocities between the low-mass Keplerian and RIAF half-and-half model is approximately $400 ~\km~\s^{-1}$.  

Increasing the radial velocities also produces narrower velocity dispersions in the center compared to Keplerian models, and models with the same total velocity magnitude have the same maximum dispersion in the center.  Away from the galaxy center, models with increased radial velocities have systematically higher velocity dispersions across the disk relative to Keplerian models, and produce peak dispersions shifted away from the center.  Some asymmetry is expected when the disk axis is rotated (rolled) with respect to the observation axis, but as radial velocity components produce more dispersion away from the galaxy center, this asymmetry is magnified.  A rotated disk can also produce a shifted velocity curve, even in the Keplerian case.  However, radial velocity components contribute most to the line-of-sight velocity orthogonally to the azimuthal components.  That is, radial velocities still contribute to the line-of-sight velocity at the slit origin, which shifts the velocity curve away from the Keplerian model, even when the disk is not rotated.

Observationally, \citet{M87gas:13} find the best-fit emissivity for M87 is two Gaussians with slight offsets from the center.  This implies that there is no drop in emissivity within $\approx 5$ pc of the galaxy center, which is the resolution of their slit.  Close to the black hole,  accretion disk temperatures should increase enough to photoionize the gas, reducing the emissivity near the center.  As mentioned above, the virial temperatures easily reach $10^8$ K at a distance of 10 pc from the galaxy center, which is just within the resolution of the observations.  The null result of reduced emissivity near the galaxy center suggests either sub-virialized flows, or that the gas clouds act as shielding for the cooler line-emitting gas inside and remain undisrupted even close to the center. 

\section{Discussion}

\subsection{Implications for the Mass of M87}

\begin{figure*}[!ht]
  \begin{center}
    \includegraphics[width=0.9\textwidth]{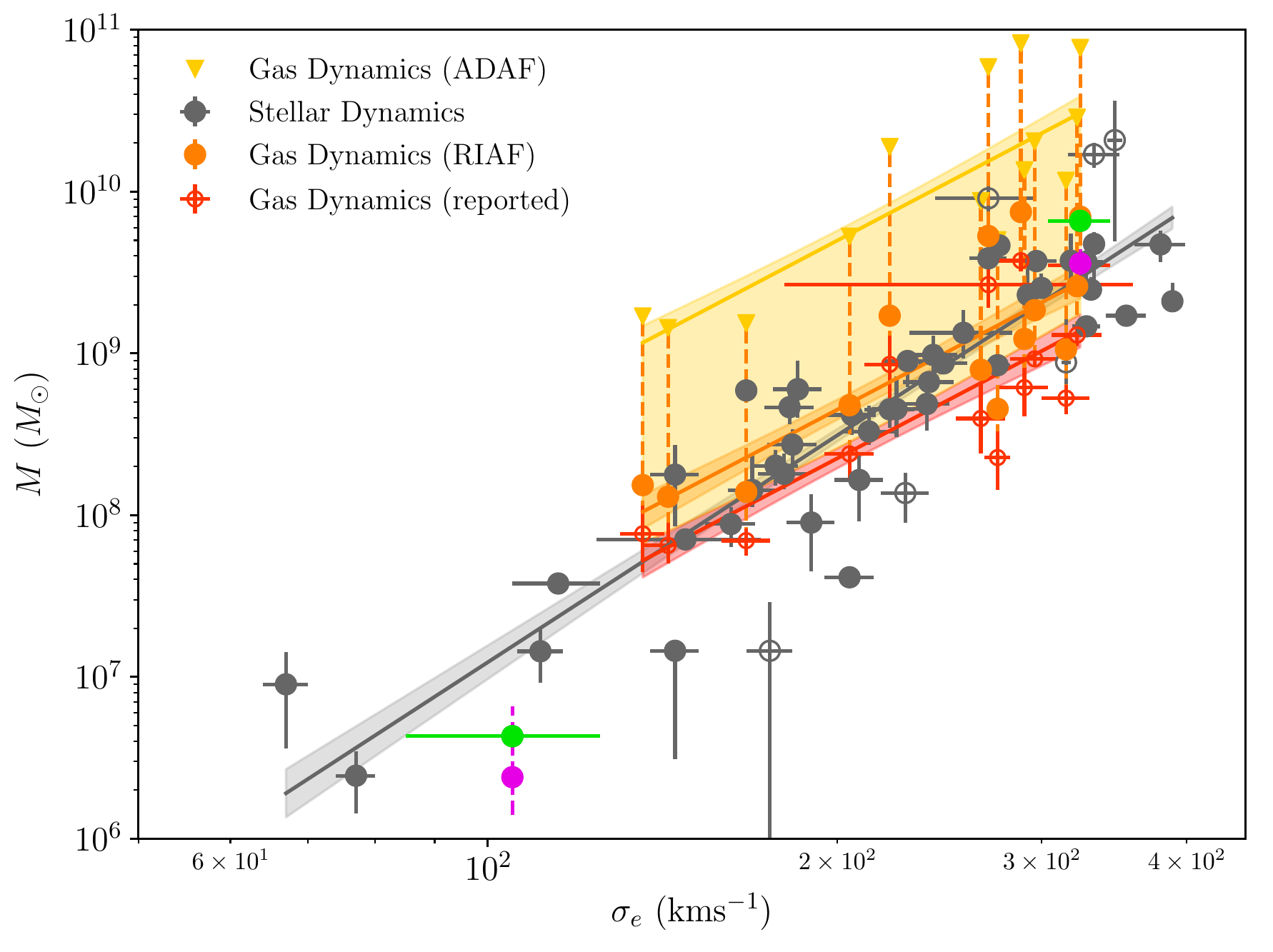}
  \end{center}
  \caption{$M$-$\sigma$ relationship for data collected in \citet{Korm-Ho:13}.  Dark gray points are SMBH mass estimates made using stellar kinematics, and open red points are reported SMBH mass estimates using ionized gas dynamics.  Orange points are twice the reported gas dynamics mass estimate, which is possible when assuming the ionized gas follows the slightly sub-Keplerian RIAF model described in this paper.  The yellow triangles are $22$ times the reported gas dynamics mass estimate, which can arise from assuming the gas is described by the very sub-Keplerian ADAF model from this paper.  The gray line is the $M$-$\sigma$ fit to the gray points, excluding the open gray points, which are the same as in \citet{Korm-Ho:13}.  The red, orange, and yellow lines are fits to the reported, RIAF, and ADAF gas models, respectively, and the gray, red, and orange shaded regions are the $2 ~\sigma$ confidence regions around each fit.  The yellow shaded region spans the breadth of the gas dynamics measurements and represents the worst case scenario for gas dynamics mass estimates.  In practice, any change in the mass estimate is likely to be much more modest, and likely lies within the orange band.  The green (stellar kinematics) and magenta (gas dynamics) points are the measurements for Sgr A* and M87.}
 \label{fig:Msig}
\end{figure*}

The key results of the velocity model described here are that models with radial motions produce asymmetric and narrower peaked dispersion profiles relative to Keplerian models, as well as shifted velocity curves away from the central slit.  Using mass and distance parameters for M87, the velocity model presented here produces a velocity dispersion of $\approx 120 ~\rm{km~s^{-1}}$  at an angular distance of $0.5''$ for the RIAF model. The data presented in Figure 4 of \citet{M87gas:13} for velocity dispersion shows observed velocity dispersions of $150\pm 50 ~\rm{km~s^{-1}}$ at $0.5''$.  The observed dispersion also shows some evidence for an asymmetric profile, especially in the 3rd and 4th slits.  In the 3rd slit, there is an $\approx 200 ~\rm{km~s^{-1}}$ difference in dispersion between $-0.1''$ and $0.1''$ along the slit.  While some asymmetry can arise from a rotation between the slit axis and disk axis, or a misaligned slit, the small rotation angle observed by \citet{M87gas:13} would preclude a difference this large.  It is important to stress that this work does not attempt to actually fit a non-Keplerian velocity model to the observed data for M87.  Rather, this work uses M87 mass and distance parameters to provide a physical reference for the effects non-Keplerian gas motions have on observed line-of-sight velocities and velocity dispersions. 

Using the non-Keplerian models presented here, it may be possible to reconcile the discrepant mass estimates for M87 while still using the observed gas velocity data.  Soon, there will be a third, independent estimate for the mass of M87 from the Event Horizon Telescope (EHT) collaboration.  The EHT is a global, millimeter-wavelength, very-long baseline interferometer (mm-VLBI) capable of resolution on the order of $10~\muas$, with the primary objective of imaging the event horizons of Sgr A* and M87 \citep{Doel_etal:08}.  Due to light-bending, the black hole should produce a shadow over the background accretion disk or jet emission with a diameter of $\approx 10 ~\rm{GM/c^2}$.  For M87, assuming a distance of $17.9 ~\rm{Mpc}$, the angular diameter of this shadow would be about $19 ~\muas$ for the mass reported in \citet{M87gas:13}, and about $36 ~\muas$ for the mass reported in \citet{M87stars:11}.  This angular diameter will be resolved by the EHT, hence the EHT should produce an unambiguous estimate of the mass of M87, independent of the gas-dynamical and stellar-dynamical measurements.  In the event the EHT produces an estimate that is higher than the lower mass estimate, the velocity models presented here can easily produce high central mass velocity and dispersion profiles qualitatively consistent with the observed gas velocity data from \citet{M87gas:13}.

\subsection{Implications for the Accretion Rate}

M87 is notably a low luminosty AGN, with a jet power estimated between $10^{42} $ and $ 10^{44} ~\erg / s^{-1}$ \citep{Bick-Begel:96,Sta-Aha-etal:06,Brom-Levin:09,Prieto:16}.  One may be concerned that the radial velocities described in this work may lead to an overestimation of the accretion rate, or a rapid depletion of the gas reservoir.  Measurements of molecular line emission place conservative estimates on the gas mass in the inner $100 ~\pc$ of M87 at about $3 \times 10^{6} ~M_{\odot}$ \citep{Tan-Beuth:08}.  However, even if this gas reservoir is being depleted via a radial velocity component, much of that infalling matter will be redirected back out via winds and other outflows before it gets to the black hole.

For RIAF accretion solutions density scales as $r^{-3/2}$ or slower, meaning $\dot{M}$ scales as $r^{0}$ to $r^{0.5}$.  This implies that as little as $0.3 \%$ of the gas in the disk gets down to the black hole \citep{Blan-Begel:99}.  The mass accretion rate onto the black hole for this disk is can be expressed as the fractional accretion rate at the edge of the disk, or 
\begin{align}
\dot{M}|_{R_{\rm in}} &= \epsilon \dot{M}|_{R_{\rm out}} = \epsilon v_r \Omega_D R^2_{\rm out} \rho
 \label{eqn:disk_mdot}
\end{align}
where $\epsilon$ is the fraction of gas that reaches the black hole, $v_r$ is the radial infall velocity, $R_{\rm in}$ is the radius where material must fall into the black hole, $R_{\rm out}$ is the edge of the accreting disk, $\Omega_D$ is the solid angle encompassing the disk, and $\rho$ is the density of the disk.  Integrating from the disk edge to the inner accretion radius allows us to express equation \ref{eqn:disk_mdot} as
\begin{align}
\dot{M}|_{R_{in}} &= \frac{\epsilon v_r}{2} \frac{M_T}{R_{out}}
\label{eqn:mdot_simple}
\end{align}
where $M_T$ is the total mass of the disk.  For a radial velocity of $0.1 v_{k}$, this implies a mass accretion rate of 
\begin{align}
\dot{M} &= \frac{\epsilon~ (0.1 ~v_{k})}{2} \left(\frac{3 \times 10^{6} ~M_{\odot}}{100 ~\pc} \right) = 2.43 \times 10^{-3} ~M_{\odot} \yr^{-1}
\label{eqn:mdot_eval}
\end{align}
where $v_{k}$ is the Keplerian velocity at $100~\pc$.  This can be converted to a luminosity via 
\begin{align}
L &= \eta \dot{M} c^{2} = 1.36 \times 10^{43} ~\erg \s^{-1}
\end{align}
where $\eta=0.1$ is a typical efficiency factor for converting accreting matter into a jet luminosity. 

For a radial velocity of $\sqrt{0.1} v_{k} ~\approx 0.3 v_{k}$, the estimated accretion luminosity is $4.1 \times 10^{43} ~\erg \s^{-1}$.  Thus for RIAF models with modest radial velocities, the estimated accretion luminosity is consistent with the observed jet luminosity.  Even large radial velocities at significant fractions of the Keplerian velocity permit acceptable accretion luminosities.  For a mass accretion rate of $\approx 10^{-3} ~M_{\odot}/\yr$, a gas disk of $10^{6} ~M_\odot$ will deplete in $\approx 1 \Gyr$, assuming no external replenishment.  

\subsection{Implications for the $M$-$\sigma$ Relation}

The correlation between the SMBH mass and bulge stellar velocity dispersions is one of the fundamental observational relationships between central black holes and their host galaxies.  For large galaxies with high ($> ~200 ~{\rm kms^{-1}}$) dispersions, and where SMBH masses have been estimated using both stellar dynamics and ionized gas dynamics, the stellar dynamics estimates produce higher black hole masses, as mentioned in Section 1.  Motivated by this discrepancy, and by the difficulty in modeling high line dispersions in ionized gas kinematics, the analysis presented in \citet{Korm-Ho:13} for the $M$-$\sigma$ relationship omits SMBH mass estimates from gas dynamics except when the mass estimates specifically attempt to incorporate corrections for the high measured dispersions.  

It is possible to apply the model presented here to the data omitted in \citet{Korm-Ho:13} and reinterpret the $M$-$\sigma$ relationship.  In Figure \ref{fig:Msig} we reproduce the right side of Figure 12 in \citet{Korm-Ho:13} to illustrate the impact of sub-Keplerian accretion flow models on the $M$-$\sigma$ relationship. The red, orange, and yellow points represent gas mass estimates assuming the reported measurements, an RIAF-like velocity field, and an ADAF velocity field, respectively.  Using the RIAF model, gas dynamics mass estimates could be increased by a factor of two, but this does not dramatically change the slope or scatter of the $M$-$\sigma$ relationship, as demonstrated by the orange fit.  Using the ADAF model, the mass estimates would increase by an order of magnitude, and increase the scatter in the relationship in the high dispersion regime, and produce some of the highest estimates for central black hole masses.  Such large black hole masses seem unreasonable, so the yellow shaded region should be interpreted as the most pessimistic region for an estimated mass shift.  In practice, each object should be individually reanalyzed with this sub-Keplerian model, and any shift in the estimated mass is expected to lie within the orange and red bands.  As $M \propto \langle v \rangle ^2$, small or moderate changes in the line-of-sight velocity from radial and sub-Keplerian motion can produce significant changes in the estimate of the central mass.  For ionized gas-dynamics, the largest systematic when estimating the SMBH mass comes from assumptions about the gas velocity model.  Presently, stellar kinematics mass estimates dominate the trend in $M$-$\sigma$, but the advent of high resolution radio interferometry from projects like ALMA  should produce many more gas dynamical SMBH mass estimates in the future.  Developing gas velocity models that can account for the sub-Keplerian systematic will be important for producing mass estimates consistent with the stellar kinematics estimates.  

The gas velocity model presented here makes it possible to simultaneously estimate the mass of the central black hole and characterize the motion of ionized gas in the gravitational influence of the central black hole.  This makes it possible to look for trends in gas motion across the mass regime, which may provide insight into why the $M$-$\sigma$ relationship saturates at high masses, relative to the $M$-$L_{\rm bulge}$ relationship.

\section{Conclusions}

By incorporating non-Keplerian velocity components, it is possible to produce velocity curves similar to those produced by circular Keplerian models in the immediate central region around the black hole, but with twice the central black hole mass.  These non-Keplerian models are distinguishable from the circular Keplerian models primarily in the velocity dispersion, producing generally higher dispersions with narrower peaks.  Outside the central region, it becomes possible to distinguish non-Keplerian velocity curves from circular Keplerian curves through a systematic shift in the center of the curve related to the magnitude of the radial velocity component.

For the specific case of M87, a non-Keplerian velocity profile (RIAF) with a radial component $\alpha=\sqrt{0.1}$ and an azimuthal component $\Omega=\sqrt{0.7}$ in units of circular Keplerian velocity, and a central black hole mass of $6.6\times 10^{9} M_{\odot}$, produces a velocity curve nearly coincident with a Keplerian velocity profile and central black hole mass of $3.5\times 10^{9} M_{\odot}$ in the central region around the black hole.  However, the peak velocity dispersion for the RIAF model is approximately $300~\km~\s^{-1}$ higher than the velocity dispersion for the purely Keplerian model.  Even outside the central region, the velocity dispersion for the RIAF model is systematically higher than the Keplerian model.  Very sub-Keplerian models, like an ADAF where the azimuthal velocity component is $\Omega=0.2$, produce qualitatively different velocity curves and dispersion profiles from Keplerian and slightly sub-Keplerian models like the RIAF.  

If we apply the model presented here to other discrepant ionized gas estimates, the incorporation of small radial velocity components would serve to increase the estimated SMBH mass.  While we did not produce a full analysis of this effect for every ionized gas estimate, there is a consistent trend where one can increase the mass estimate by incorporating radial and sub-Keplerian velocity components.  Assuming modest amounts of radial motion can increase the mass estimate by a factor of two, while not significantly affecting the $M$-$\sigma$ relationship at the high $\sigma_e$ range.  Assuming substantially sub-Keplerian motion, like produced in the ADAF model, can increase the mass estimate by well over an order of magnitude, and produce unreasonably high black hole mass estimates.  The choice of how to model the dynamics of ionized gas around central black holes is a significant systematic when estimating the black hole mass, and in extreme cases can lead to order-of-magnitude differences in the estimated black hole mass.  

While the discussion presented here is focused on the particular case of the SMBH at the center of M87, the model we present is applicable to gas-dynamical SMBH mass estimates in general, as discussed in Section 4.2.  By modifying the magnitude of the radial and azimuthal velocity components, as well as the fraction of virial dispersion contributing to the total line dispersion, it will be possible not only to better model real observations of nuclear line emission, but also simultaneously characterize the gas motions around the central black hole.

\acknowledgments
The authors would like to thank the anonymous referee and Aaron Barth for helpful comments.
This work was supported in part by the Perimeter Institute for Theoretical Physics. Research at Perimeter Institute is supported by the Government of Canada through the Department of Innovation, Science and Economic Development Canada and by the Province of Ontario through the Ministry of Research, Innovation and Science.
A.E.B. thanks the Delaney Family for their generous financial support via the Delaney Family John A. Wheeler Chair at Perimeter Institute.
B.J. and A.E.B receive additional financial support from the Natural Sciences and Engineering Research Council of Canada through a Discovery Grant.
B.R.M thanks the Natural Science and Engineering Research Council for generous support.

\bibliographystyle{apj}
\bibliography{m87}

\begin{thebibliography}{}
\expandafter\ifx\csname natexlab\endcsname\relax\def\natexlab#1{#1}\fi

\bibitem[{{Barth} {et~al.}(2001){Barth}, {Sarzi}, {Ho}, {Rix}, {Shields},
  {Filippenko}, {Rudnick}, \& {Sargent}}]{Barth-Sarz:01}
{Barth}, A.~J., {Sarzi}, M., {Ho}, L.~C., {et~al.} 2001, in Astronomical
  Society of the Pacific Conference Series, Vol. 249, The Central Kiloparsec of
  Starbursts and AGN: The La Palma Connection, ed. J.~H. {Knapen}, J.~E.
  {Beckman}, I.~{Shlosman}, \& T.~J. {Mahoney}, 370

\bibitem[{{Becklin} {et~al.}(1982){Becklin}, {Gatley}, \& {Werner}}]{Beck:82}
{Becklin}, E.~E., {Gatley}, I., \& {Werner}, M.~W. 1982, \apj, 258, 135

\bibitem[{{Bicknell} \& {Begelman}(1996)}]{Bick-Begel:96}
{Bicknell}, G.~V., \& {Begelman}, M.~C. 1996, \apj, 467, 597

\bibitem[{{Bird} {et~al.}(2010){Bird}, {Harris}, {Blakeslee}, \&
  {Flynn}}]{Bird-Harris:10}
{Bird}, S., {Harris}, W.~E., {Blakeslee}, J.~P., \& {Flynn}, C. 2010, \aap,
  524, A71

\bibitem[{{Blandford} \& {Begelman}(1999)}]{Blan-Begel:99}
{Blandford}, R.~D., \& {Begelman}, M.~C. 1999, \mnras, 303, L1

\bibitem[{{Blandford} \& {McKee}(1982)}]{Bland-McKee:82}
{Blandford}, R.~D., \& {McKee}, C.~F. 1982, \apj, 255, 419

\bibitem[{{Boizelle} {et~al.}(2019){Boizelle}, {Barth}, {Walsh}, {Buote},
  {Baker}, {Darling}, \& {Ho}}]{Boiz-Barth-Walsh:19}
{Boizelle}, B.~D., {Barth}, A.~J., {Walsh}, J.~L., {et~al.} 2019, arXiv
  e-prints, arXiv:1906.06267

\bibitem[{{Bromberg} \& {Levinson}(2009)}]{Brom-Levin:09}
{Bromberg}, O., \& {Levinson}, A. 2009, \apj, 699, 1274

\bibitem[{{Burkert} \& {Tremaine}(2010)}]{Burk-Trem:10}
{Burkert}, A., \& {Tremaine}, S. 2010, \apj, 720, 516

\bibitem[{{Cantiello} {et~al.}(2018){Cantiello}, {Blakeslee}, {Ferrarese},
  {C{\^o}t{\'e}}, {Roediger}, {Raimondo}, {Peng}, {Gwyn}, {Durrell}, \&
  {Cuillandre}}]{Cantiello:18}
{Cantiello}, M., {Blakeslee}, J.~P., {Ferrarese}, L., {et~al.} 2018, \apj, 856,
  126

\bibitem[{{Cappellari} {et~al.}(2009){Cappellari}, {Neumayer}, {Reunanen}, {van
  der Werf}, {de Zeeuw}, \& {Rix}}]{Cap-CenAStars:07}
{Cappellari}, M., {Neumayer}, N., {Reunanen}, J., {et~al.} 2009, \mnras, 394,
  660

\bibitem[{{Chan} \& {Krolik}(2017)}]{Chan-Krol:17}
{Chan}, C.-H., \& {Krolik}, J.~H. 2017, \apj, 843, 58

\bibitem[{{Davies} {et~al.}(2006){Davies}, {Thomas}, {Genzel}, {M{\"u}ller
  S{\'a}nchez}, {Tacconi}, {Sternberg}, {Eisenhauer}, {Abuter}, {Saglia}, \&
  {Bender}}]{Dav-Thom:06}
{Davies}, R.~I., {Thomas}, J., {Genzel}, R., {et~al.} 2006, \apj, 646, 754

\bibitem[{{Davis} {et~al.}(2017){Davis}, {Bureau}, {Onishi}, {Cappellari},
  {Iguchi}, \& {Sarzi}}]{Davis_etal:17}
{Davis}, T.~A., {Bureau}, M., {Onishi}, K., {et~al.} 2017, \mnras, 468, 4675

\bibitem[{{de Francesco} {et~al.}(2006){de Francesco}, {Capetti}, \&
  {Marconi}}]{deFranc:06}
{de Francesco}, G., {Capetti}, A., \& {Marconi}, A. 2006, \aap, 460, 439

\bibitem[{{Doeleman} {et~al.}(2008)}]{Doel_etal:08}
{Doeleman}, S.~S., {et~al.} 2008, \nat, 455, 78

\bibitem[{{Eckart} {et~al.}(2002){Eckart}, {Genzel}, {Ott}, \&
  {Sch{\"o}del}}]{Eckart:02}
{Eckart}, A., {Genzel}, R., {Ott}, T., \& {Sch{\"o}del}, R. 2002, \mnras, 331,
  917

\bibitem[{{Fabian}(2012)}]{Fab:12}
{Fabian}, A.~C. 2012, \araa, 50, 455

\bibitem[{{Ferrarese}(2002)}]{Ferr:02}
{Ferrarese}, L. 2002, \apj, 578, 90

\bibitem[{{Ferrarese} \& {Ford}(2005)}]{Ferr-Ford:05}
{Ferrarese}, L., \& {Ford}, H. 2005, \ssr, 116, 523

\bibitem[{{Ferrarese} \& {Merritt}(2000)}]{Ferr-Merr:00}
{Ferrarese}, L., \& {Merritt}, D. 2000, \apjl, 539, L9

\bibitem[{{Gebhardt} {et~al.}(2011){Gebhardt}, {Adams}, {Richstone}, {Lauer},
  {Faber}, {G{\"u}ltekin}, {Murphy}, \& {Tremaine}}]{M87stars:11}
{Gebhardt}, K., {Adams}, J., {Richstone}, D., {et~al.} 2011, \apj, 729, 119

\bibitem[{{Gebhardt} \& {Richstone}(2000)}]{Gebh-Rich:00}
{Gebhardt}, K., \& {Richstone}, D.~O. 2000, in Bulletin of the American
  Astronomical Society, Vol.~32, American Astronomical Society Meeting
  Abstracts \#196, 700

\bibitem[{{Gebhardt} {et~al.}(2000){Gebhardt}, {Bender}, {Bower}, {Dressler},
  {Faber}, {Filippenko}, {Green}, {Grillmair}, {Ho}, {Kormendy}, {Lauer},
  {Magorrian}, {Pinkney}, {Richstone}, \& {Tremaine}}]{Geb-Msigma:00}
{Gebhardt}, K., {Bender}, R., {Bower}, G., {et~al.} 2000, \apjl, 539, L13

\bibitem[{{Genzel} {et~al.}(2010){Genzel}, {Eisenhauer}, \&
  {Gillessen}}]{Genz:10}
{Genzel}, R., {Eisenhauer}, F., \& {Gillessen}, S. 2010, Reviews of Modern
  Physics, 82, 3121

\bibitem[{{Ghez} {et~al.}(2009){Ghez}, {Morris}, {Lu}, {Weinberg}, {Matthews},
  {Alexander}, {Armitage}, {Becklin}, {Brown}, {Campbell}, {Do}, {Eckart},
  {Genzel}, {Gould}, {Hansen}, {Ho}, {Lo}, {Loeb}, {Melia}, {Merritt},
  {Milosavljevic}, {Perets}, {Rasio}, {Reid}, {Salim}, {Sch{\"o}del}, \&
  {Yelda}}]{Ghez:09}
{Ghez}, A., {Morris}, M., {Lu}, J., {et~al.} 2009, in Astronomy, Vol. 2010,
  astro2010: The Astronomy and Astrophysics Decadal Survey

\bibitem[{{Gillessen} {et~al.}(2009{\natexlab{a}}){Gillessen}, {Eisenhauer},
  {Fritz}, {Bartko}, {Dodds-Eden}, {Pfuhl}, {Ott}, \& {Genzel}}]{GillS2:09}
{Gillessen}, S., {Eisenhauer}, F., {Fritz}, T.~K., {et~al.} 2009{\natexlab{a}},
  \apjl, 707, L114

\bibitem[{{Gillessen} {et~al.}(2009{\natexlab{b}}){Gillessen}, {Eisenhauer},
  {Trippe}, {Alexander}, {Genzel}, {Martins}, \& {Ott}}]{Gil-SgrAStars:09}
{Gillessen}, S., {Eisenhauer}, F., {Trippe}, S., {et~al.} 2009{\natexlab{b}},
  \apj, 692, 1075

\bibitem[{{Harris} \& {Harris}(2011)}]{HandH:11}
{Harris}, G.~L.~H., \& {Harris}, W.~E. 2011, \mnras, 410, 2347

\bibitem[{{Imanishi} {et~al.}(2018){Imanishi}, {Nakanishi}, \&
  {Izumi}}]{Iman:18}
{Imanishi}, M., {Nakanishi}, K., \& {Izumi}, T. 2018, \apj, 856, 143

\bibitem[{{Irons} {et~al.}(2012){Irons}, {Lacy}, \&
  {Richter}}]{IronsLacy-SgrASpiral:12}
{Irons}, W.~T., {Lacy}, J.~H., \& {Richter}, M.~J. 2012, \apj, 755, 90

\bibitem[{{Kormendy} \& {Bender}(2011)}]{Korm-Bend:11}
{Kormendy}, J., \& {Bender}, R. 2011, \nat, 469, 377

\bibitem[{{Kormendy} \& {Ho}(2013)}]{Korm-Ho:13}
{Kormendy}, J., \& {Ho}, L.~C. 2013, \araa, 51, 511

\bibitem[{{Kormendy} \& {Richstone}(1995)}]{Korm-Rich:95}
{Kormendy}, J., \& {Richstone}, D. 1995, \araa, 33, 581

\bibitem[{{Lacy} {et~al.}(1979){Lacy}, {Baas}, {Townes}, \&
  {Geballe}}]{Lacy:79}
{Lacy}, J.~H., {Baas}, F., {Townes}, C.~H., \& {Geballe}, T.~R. 1979, \apjl,
  227, L17

\bibitem[{{Lacy} {et~al.}(1980){Lacy}, {Townes}, {Geballe}, \&
  {Hollenbach}}]{Lacy:80}
{Lacy}, J.~H., {Townes}, C.~H., {Geballe}, T.~R., \& {Hollenbach}, D.~J. 1980,
  \apj, 241, 132

\bibitem[{{Lynden-Bell}(1978)}]{Lyn:1978}
{Lynden-Bell}, D. 1978, \physscr, 17, 185

\bibitem[{{Macchetto} {et~al.}(1997){Macchetto}, {Marconi}, {Axon}, {Capetti},
  {Sparks}, \& {Crane}}]{Macc-M87gas:97}
{Macchetto}, F., {Marconi}, A., {Axon}, D.~J., {et~al.} 1997, \apj, 489, 579

\bibitem[{{Marconi} \& {Hunt}(2003)}]{Marc-Hunt:03}
{Marconi}, A., \& {Hunt}, L.~K. 2003, \apjl, 589, L21

\bibitem[{{Mazzucchelli} {et~al.}(2017){Mazzucchelli}, {Ba{\~n}ados},
  {Venemans}, {Decarli}, {Farina}, {Walter}, {Eilers}, {Rix}, {Simcoe},
  {Stern}, {Fan}, {Schlafly}, {De Rosa}, {Hennawi}, {Chambers}, {Greiner},
  {Burgett}, {Draper}, {Kaiser}, {Kudritzki}, {Magnier}, {Metcalfe}, {Waters},
  \& {Wainscoat}}]{Maz-Ban:17}
{Mazzucchelli}, C., {Ba{\~n}ados}, E., {Venemans}, B.~P., {et~al.} 2017, \apj,
  849, 91

\bibitem[{{McNamara} \& {Nulsen}(2012)}]{McNam-Nul:12}
{McNamara}, B.~R., \& {Nulsen}, P.~E.~J. 2012, New Journal of Physics, 14,
  055023

\bibitem[{{Montero-Casta{\~n}o} {et~al.}(2009){Montero-Casta{\~n}o},
  {Herrnstein}, \& {Ho}}]{Mont-Hern-Ho:09}
{Montero-Casta{\~n}o}, M., {Herrnstein}, R.~M., \& {Ho}, P.~T.~P. 2009, \apj,
  695, 1477

\bibitem[{{Narayan} {et~al.}(1998){Narayan}, {Mahadevan}, \&
  {Quataert}}]{Nar-Mah-Quat:98}
{Narayan}, R., {Mahadevan}, R., \& {Quataert}, E. 1998, in Theory of Black Hole
  Accretion Disks, ed. M.~A. {Abramowicz}, G.~{Bj{\"o}rnsson}, \& J.~E.
  {Pringle}, 148--182

\bibitem[{{Narayan} \& {Yi}(1994)}]{Nar-Yi-ADAF:94}
{Narayan}, R., \& {Yi}, I. 1994, \apjl, 428, L13

\bibitem[{{Narayan} \& {Yi}(1995)}]{Nar-Yi-ADAF:95}
---. 1995, \apj, 444, 231

\bibitem[{{Neumayer} {et~al.}(2007){Neumayer}, {Cappellari}, {Reunanen}, {Rix},
  {van der Werf}, {de Zeeuw}, \& {Davies}}]{Neu-CenAGas:07}
{Neumayer}, N., {Cappellari}, M., {Reunanen}, J., {et~al.} 2007, \apj, 671,
  1329

\bibitem[{{Padmanabhan}(2000)}]{Padm-book:00}
{Padmanabhan}, T. 2000, {Theoretical Astrophysics - Volume 1, Astrophysical
  Processes}, 622, doi:10.2277/0521562406

\bibitem[{{Park} \& {Ostriker}(1999)}]{Park-Ostr:99}
{Park}, M.-G., \& {Ostriker}, J.~P. 1999, \apj, 527, 247

\bibitem[{{Pastorini} {et~al.}(2007){Pastorini}, {Marconi}, {Capetti}, {Axon},
  {Alonso-Herrero}, {Atkinson}, {Batcheldor}, {Carollo}, {Collett}, {Dressel},
  {Hughes}, {Macchetto}, {Maciejewski}, {Sparks}, \& {van der
  Marel}}]{Past-Marc:07}
{Pastorini}, G., {Marconi}, A., {Capetti}, A., {et~al.} 2007, \aap, 469, 405

\bibitem[{{Peterson} \& {Bentz}(2011)}]{PeterReverb:11}
{Peterson}, B.~M., \& {Bentz}, M.~C. 2011, {Black-hole masses from
  reverberation mapping}, ed. M.~{Livio} \& A.~M. {Koekemoer}, 100--111

\bibitem[{{Prieto} {et~al.}(2016){Prieto}, {Fern{\'a}ndez-Ontiveros},
  {Markoff}, {Espada}, \& {Gonz{\'a}lez-Mart{\'{\i}}n}}]{Prieto:16}
{Prieto}, M.~A., {Fern{\'a}ndez-Ontiveros}, J.~A., {Markoff}, S., {Espada}, D.,
  \& {Gonz{\'a}lez-Mart{\'{\i}}n}, O. 2016, \mnras, 457, 3801

\bibitem[{{Quataert} \& {Narayan}(1999)}]{Quat-Nar:99}
{Quataert}, E., \& {Narayan}, R. 1999, \apj, 516, 399

\bibitem[{{Sch{\"o}del} {et~al.}(2002){Sch{\"o}del}, {Ott}, {Genzel},
  {Hofmann}, {Lehnert}, {Eckart}, {Mouawad}, {Alexander}, {Reid}, {Lenzen},
  {Hartung}, {Lacombe}, {Rouan}, {Gendron}, {Rousset}, {Lagrange}, {Brandner},
  {Ageorges}, {Lidman}, {Moorwood}, {Spyromilio}, {Hubin}, \&
  {Menten}}]{Scho:02}
{Sch{\"o}del}, R., {Ott}, T., {Genzel}, R., {et~al.} 2002, \nat, 419, 694

\bibitem[{{Shen}(2013)}]{ShenReverb:13}
{Shen}, Y. 2013, Bulletin of the Astronomical Society of India, 41, 61

\bibitem[{{Stawarz} {et~al.}(2006){Stawarz}, {Aharonian}, {Kataoka},
  {Ostrowski}, {Siemiginowska}, \& {Sikora}}]{Sta-Aha-etal:06}
{Stawarz}, {\L}., {Aharonian}, F., {Kataoka}, J., {et~al.} 2006, \mnras, 370,
  981

\bibitem[{{Tan} {et~al.}(2008){Tan}, {Beuther}, {Walter}, \&
  {Blackman}}]{Tan-Beuth:08}
{Tan}, J.~C., {Beuther}, H., {Walter}, F., \& {Blackman}, E.~G. 2008, \apj,
  689, 775

\bibitem[{{Verdoes Kleijn} {et~al.}(2002){Verdoes Kleijn}, {van der Marel}, {de
  Zeeuw}, {Noel-Storr}, \& {Baum}}]{Verdoes:02}
{Verdoes Kleijn}, G.~A., {van der Marel}, R.~P., {de Zeeuw}, P.~T.,
  {Noel-Storr}, J., \& {Baum}, S.~A. 2002, \aj, 124, 2524

\bibitem[{{Walsh} {et~al.}(2013){Walsh}, {Barth}, {Ho}, \& {Sarzi}}]{M87gas:13}
{Walsh}, J.~L., {Barth}, A.~J., {Ho}, L.~C., \& {Sarzi}, M. 2013, \apj, 770, 86

\bibitem[{{Walsh} {et~al.}(2010){Walsh}, {Barth}, \& {Sarzi}}]{Walsh:10}
{Walsh}, J.~L., {Barth}, A.~J., \& {Sarzi}, M. 2010, \apj, 721, 762

\bibitem[{{Walsh} {et~al.}(2012){Walsh}, {van den Bosch}, {Barth}, \&
  {Sarzi}}]{Walsh:12}
{Walsh}, J.~L., {van den Bosch}, R.~C.~E., {Barth}, A.~J., \& {Sarzi}, M. 2012,
  \apj, 753, 79

\end{thebibliography}
\end{document}